\documentclass{article}
\usepackage{arxiv}

\usepackage[utf8]{inputenc} 
\usepackage[T1]{fontenc}    
\usepackage{hyperref}       
\usepackage{url}            
\usepackage{booktabs}       
\usepackage{amsfonts}       
\usepackage{amsmath}        
\usepackage{nicefrac}       
\usepackage{microtype}      
\usepackage{lipsum}		
\usepackage{graphicx}
\usepackage{natbib}
\usepackage{doi}

\title{Holistic Information Theory of Spatial Remote Sensing Imaging}

\author{ \href{https://orcid.org/0009-0009-8305-4285}{\includegraphics[scale=0.06]{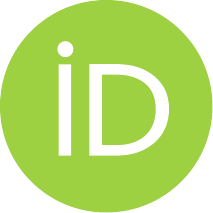}\hspace{1mm}Jianan PAN}\thanks{corresponding email: jnanpan@outlook.com} \\
	Advanced Optics Research Laboratory \\
    Chang Guang Satellite Technology Company Ltd. \\
	\texttt{jnanpan@outlook.com} \\
	\And
	\href{https://orcid.org/0000-0003-1023-2203}{\includegraphics[scale=0.06]{orcid.pdf}\hspace{1mm}Junbo HAO} \\
	Advanced Optics Research Laboratory \\
    Chang Guang Satellite Technology Company Ltd. \\
	\texttt{junbohao0413@163.com} \\
	\And
	{\hspace{1mm}Qixiang GAO} \\
	Advanced Optics Research Laboratory \\
    Chang Guang Satellite Technology Company Ltd. \\
	\texttt{gqx17@tsinghua.org.cn} \\
    \And
	\href{https://orcid.org/0000-0001-8819-4687}{\includegraphics[scale=0.06]{orcid.pdf}\hspace{1mm}Xing ZHONG} \\
	Advanced Optics Research Laboratory \\
    Chang Guang Satellite Technology Company Ltd. \\
	\texttt{ciomper@163.com} \\
}

\renewcommand{\shorttitle}{\textit{arXiv} Template}

\hypersetup{
pdftitle={A template for the arxiv style},
pdfsubject={q-bio.NC, q-bio.QM},
pdfauthor={David S.~Hippocampus, Elias D.~Striatum},
pdfkeywords={First keyword, Second keyword, More},
}

\begin{document}
\maketitle

\begin{abstract}
To address the non-optimal global design caused by the independent optimization of optical lenses, photodetectors, 
  and computational processing subsystems in traditional remote sensing imaging system design, this paper 
  proposes a holistic information theory for spatial remote sensing imaging. This theory integrates 
  the optoelectronic imaging hardware front end and computational reconstruction back end into a unified framework. 
  It establishes a complete spatial imaging chain information transfer model with the objective of obtaining 
  the required effective information. The paper innovatively proposes a quantifiable Modulation Transfer 
  Function (MTF)-Signal-to-Noise Ratio (SNR) product criterion. 
  It demonstrates that the system information transmission ability is determined by the product of MTF and SNR, 
  and that these parameters can compensate for each other to achieve equivalent information transfer. 
  Validation through a high-resolution Earth observation system case shows that under consistent reconstruction 
  mean square error conditions, increasing time delay integration stages reduces optical aperture size and 
  significantly lowers primary mirror mass. Simulations and physical experiments further indicate that 
  by increasing integration time, low-resolution optical systems achieve reconstructed fidelity comparable 
  to high-resolution systems. This verifies that small-aperture optical systems can achieve equivalent imaging 
  performance by enhancing SNR. This theory has been successfully applied in the design of the Jilin-1 
  satellite constellation, providing a new paradigm for low-cost high-resolution remote sensing systems.
\end{abstract}


\section{Introduction}\label{1}
Remote sensing imaging typically refers to using aircraft or satellites equipped with
space optical instruments to measure the characteristics of objects on Earth
non-contactly, relying on the propagation of light signals \cite{schowengerdt2006remote}.
Remote sensing imaging provides repeated and consistent observations of the Earth, 
which are crucial for monitoring environments and human activities. 
It has a wide range of applications, including transportation, agriculture, 
and marine environments \cite{guo2020manual}.
In recent years, with the continuous advancements in satellite remote sensing manufacturing and 
microchip processing technologies, the resolution of remote sensing data has consistently 
improved across spatial and spectral dimensions \cite{zhang2022progress}. 
The enhanced spatial resolution significantly increases the value of remote sensing 
images.
\par
Based on the concept of diffraction limitation, the resolution of an optical system is
determined by its aperture size, which means that achieving higher spatial resolution 
in space optical systems requires larger aperture sizes.  This leads to satellite platforms 
having greater mass and volume, resulting in increased manufacturing costs.
Several approaches have been proposed to alleviate the increasing deployment costs 
associated with pursuing high spatial resolution in optical systems by 
enlarging the aperture. The James Webb Space Telescope \cite{JWST} employs segmented 
mirror technology \cite{mesrine2017high}, which addresses the challenges of manufacturing large 
aperture mirrors and the inability of rockets to carry such large mirrors by 
producing in blocks and deploying in folds.
Lightweighting of optical-mechanical structures is researched \cite{33Lightweight},
like SPICA \cite{34spica} utilizing carbon fiber-reinforced plastics 
to support specific structures, forming a lightweight architecture.
Other advanced research which may be applied in space imaging is ongoing.
The imaging performance and deployment costs 
of various systems using a sparse array of mirror elements as the 
primary mirror of reflective optical systems are explored by \cite{40size_reduce}, based on the concept of 
optical sparse aperture \cite{35OSA}.
The theoretical and manufacturing of large-aperture metalenses are 
studied by \cite{7HAO} and \cite{41large}.
FalconSAT-7 \cite{8FalconSAT} employs diffractive elements as a collecting aperture, 
significantly reducing the mass and volume of the optical system.
\par
Due to the diffraction of electromagnetic waves during propagation in space, 
the spatial positions of physical parameters become distorted when non-contact 
methods are employed to measure optical information from targets. 
Traditional imaging instruments utilize focusing optical elements to converge 
wavefronts onto photodetectors, thereby establishing a spatial mapping between 
objects and images to achieve optical information measurement. 
In recent years, with continuous advancements in information science, 
computational imaging \cite{CI} has been widely applied in optoelectronic 
imaging fields. 
Distinct from conventional optical systems where spatial mapping relies entirely 
on optical lenses, computational imaging systems aim to implement spatial mapping 
through co-designed front-end and back-end strategies. 
This approach leverages the advantages of information technology while 
circumventing bottlenecks in optical system design and manufacturing. 
The first widespread application of computational-optical joint imaging methods 
can be traced to the pioneering work in \cite{CISR}. After the proposal of the
depth-of-field extension technology \cite{ED}, which has significant 
practical applications, this field has gradually gained attention.
Modulation in front-end imaging systems often involves hardware modifications. 
Consequently, fields like microscopy that permit active hardware modulation 
particularly benefit from computational imaging techniques, 
such as Fourier ptychography \cite{FP} and lensless imaging \cite{LL}. 
In contrast, remote sensing spatial imaging systems face limited adoption 
of computational optical technology due to high hardware iteration costs, 
difficulties in modulating front-end imaging hardware, and engineering inertia. 
Thus, imaging systems and image processing systems in remote sensing often 
remain separately designed.
The capability of large-scale satellite constellations for repeat Earth observation, 
combined with the macroscopic nature of remote sensing applications, 
results in substantially lower temporal resolution requirements compared to 
microscopy, which operates at sub-second levels \cite{subS}. 
Remote sensing typically maintains daily-level temporal resolution \cite{RSS}, 
allowing high-performance image processing techniques to play a more prominent role. 
Therefore, effectively leveraging computational imaging to organically integrate 
optoelectronic hardware with image algorithms and achieving 
optimal imaging performance under constrained engineering costs 
represents a crucial developmental imperative for remote sensing 
imaging technology.
\par

\begin{table}[htbp]
  \caption{Nomenclature}\label{Nomenclature}
  \centering
  \begin{tabular}{lll}
  \toprule
  Symbol & Definition [Unit] \\ 
  \midrule
  $U$ & Light field [V/m] \\
  $(x, y)$ & 2D spatial position coordinates [m] \\
  $(u, v)$ & 2D spatial frequency coordinates [m$^{-1}$] \\ 
  $z$ & Axial coordinates [m] \\
  $\mathbb R$ & Field of Real Numbers \\
  $k$ & Wave vector [m$^{-1}$] \\
  $\lambda$ & Wavelength [m] \\
  $I$ & Optical field intensity [W/m$^2$] \\
  $\epsilon_0$ & Vacuum permittivity [F$\cdot$ m$^2$] \\
  $c$ & Lightspeed [m/s] \\
  $S_x$ & Field of View [m$^2$] \\
  $S_f$ & Optical bandwidth [m$^{-2}$]\\
  $H_{\mathrm{orbit}}$ & Satellite orbit altitude [m]\\
  $\mathcal F$ & 2D Fourier transform operator\\
  SBP & Space-bandwidth product\\
  $f$ & Focal length [m]\\
  $Z$ & Pixel number\\
  $\Delta$ & Pixel size\\
  $F/\#$ & F-number\\
  $p$ & Photon number\\
  $\eta$ & Quantum efficiency\\
  $t$ & Integration time [s]\\
  $h$ & Planck constant [J$\cdot$s]\\
  $\mathrm{DN}$ & Digital number \\
  $K_\mathrm{gain}$ & Mapping gain from photon to digital number \\
  $\mathrm{rect}$ & Window function \\
  $\mathrm{comb}$ & Comb function \\
  $(m, n)$ & Discrete coordinates \\
  $F$ & Image spectrum \\
  $C$ & Channel capacity [bit / m$^{2}$]\\
  $B$ & Channel bandwidth [m$^{-2}$] \\
  $I_\mathrm{info}$ & Maximal mutual information [bit] \\
  SNR & Signal-to-noise ratio \\
  MTF & Modulation transfer function \\
  $H$ & Transfer function \\
  $N$ & Noise spectrum \\
  $H_\mathrm{info}$ & Information entropy [bit] \\
  $n_p$ & Photon shot noise \\
  $E$ & Expectation \\
  $D$ & Variance \\
  $X \times Y$ & The set of spatial position coordinates \\
  $F_u \times F_v$ & The set of spatial frequency coordinates \\
  $\varepsilon$ & Reconstruction error \\
  NPM & Noise power magnification \\ 
  $M$ & TDI stage \\
  \bottomrule
  \end{tabular}
  \end{table}

As an information transmission system, the ultimate design goal of the
remote sensing spatial imaging system is to obtain predetermined information content 
at the lowest possible cost. 
Designers typically decompose this goal into several subsystem-level objectives:
1) The optical system is designed, manufactured, and aligned meticulously to achieve 
optimal imaging performance with a finite aperture; 
2) Semiconductor devices and electronic circuits are optimized through design 
and process technology to minimize image noise contamination to required levels;
3) Degradation compensations including radiometric correction, sharpening, and 
denoising are implemented to the greatest extent possible in the image processing 
subsystem. During optical payload manufacturing and remote sensing data production, 
optimization of these local components is typically performed without 
considering constraints from other subsystems, or only accounting for 
immediately adjacent ones. For example, MTF design targets for optical lenses
solely focus on meeting specifications at the Nyquist frequency determined 
by detectors \cite{FNp}, and MTF compensation (MTFc) algorithms 
for mitigating image blur and noise 
are heuristic or estimated usually \cite{MTFcHJ1A, MTFcCBERS}.
Although all subsystems are designed to meet performance requirements at minimal cost, 
combinations of such locally optimal solutions often fail to achieve global optimality.
For example, due to the heuristic MTFc algorithm, the improvement in optical system 
performance does not guarantee enhanced quality in the final processed images.
This demonstrates that engineering design generally lacks the 
property of optimal substructure \cite{ComOpt}.
Fortunately, because the physical models and stochastic processes 
governing the imaging system are nearly well-defined, integrating this 
multidisciplinary design optimization into a monolithic architecture 
\cite{MulOpt} is feasible.
For example, the instrument design principles proposed by the French 
National Centre for Space Studies state that the value of the MTF-SNR 
product is an important instrument design objective \cite{CNES1, CNES2}. 
and it is widely used in the satellite industry.
Additionally, 
with the development of AI technology in recent years, both MTF 
and SNR are core metrics for evaluating the effectiveness of intelligent 
image processing algorithms \cite{AI_MTF_SNR, AI_MTF}.
However, theoretical and quantifiable research has not yet been implemented.
Therefore, this paper treats the imaging front-end of optical payload cameras and 
the computational back-end as a unified system.
Guided by the objective of obtaining the required effective information, 
we conduct an optimization analysis for the whole imaging system design.
This process involves transformations among optical information, 
electronic information, and digital information.
Consequently, we propose a method called holistic information 
theory for remote 
sensing optical design based on this principle.
The nomenclature of this section is described in Table \ref{Nomenclature}.

\section{Holistic Information Analysis under the Spatial Imaging Chain}\label{2}

For traditional spatial imaging chains, 
the physical models and stochastic processes are nearly well-defined.
However, conventional design methods struggle to ensure globally 
optimal solutions when combining different subsystems, 
consequently relying on heuristically proposed principles \cite{p1}.
This section adopts an information-theoretic perspective to examine 
information transfer throughout the spatial imaging chain, 
while further investigating how subsystem selections impact 
systemic information transfer, thus establishing a physical 
foundation for quantifying holistic information.

\subsection{The Propagation of Optical Information}\label{2_1}

The target information source of remote sensing imaging systems is the radiance 
of sunlight reflected from ground objects. At any given moment, the acquired 
information constitutes a two-dimensional light field at the conjugate plane 
$z_{obj}$ of the detector array, which can be regarded as a linear superposition 
of plane waves with different propagation directions as
\begin{equation}\label{Uobj}
  \begin{aligned}
    U(x, y, z_\mathrm{obj}) = \iint_{\mathbb{R}^2} A(u, v, z_\mathrm{obj})
    \mathrm{exp}[j2\pi(ux + vy)]dudv,
  \end{aligned}
\end{equation}
where $A(u, v, z_{obj})$ is the angular spectrum in spatial frequency $(u, v)$
and conjugate plane $z_\mathrm{obj}$, $U(x, y, z_\mathrm{obj})$ is 
the complex amplitude in 
spatial position $(x, y)$ and conjugate plane $z_\mathrm{obj}$.
the free space propagation of the angular spectrum satisfies
\begin{equation}\label{Helmholtz}
  \begin{aligned}
    A(u, v, z_\mathrm{obj} + \Delta z) = A(u, v, z_\mathrm{obj})
    \mathrm{exp}(jk\Delta z\sqrt{
      1 - \lambda^2 (u^2 + v^2) 
    })
  \end{aligned}
\end{equation}
based on the Helmholtz equation.
It is evident that the light field itself is bandlimited to $1/\lambda$ 
when evanescent waves are neglected. Furthermore, although the light field 
is a spatially continuous physical quantity, the spatially bounded domain 
over which we can capture it inevitably constrains the information content 
carried by the observable light field. 
To quantify the information-carrying capacity of the light field, 
the space-bandwidth product (SBP) is introduced to measure 
the information content of the light field \cite{SBP}. 
Taking the information carried by the light field intensity 
$I(x, y) = |U(x, y)|^2/2\epsilon_0 c$ as an example, there is
\begin{equation}\label{SBP}
  \begin{aligned}
    \mathrm{SBP} = S_xS_f.
  \end{aligned}
\end{equation}
The light field intensity can be represented by this number of independent
orthogonal signal components.
Moreover, the phase, polarization states, and others of the light field 
can also carry information.
\par
Returning to the spatial imaging chain, 
the imaging of the light field through the optical system 
is affected by the diffraction limit caused by a finite 
aperture and aberration in engineering implementation. 
In remote sensing for Earth observation, 
the light field reflected by ground objects, denoted as $U_{\mathrm{obj}}$, 
undergoes far-field diffraction over the satellite orbit altitude $H_\mathrm{orbit}$, 
reaches the satellite optical payload, and at this point, 
the light field $U_{\mathrm{in}}$ is proportional to the spectral distribution 
of $U_{\mathrm{obj}}$ as
\begin{equation}\label{FDD}
  \begin{aligned}
    U_{\mathrm{in}}(x_{\mathrm{opt}}, y_{\mathrm{opt}}) \propto
    \mathcal{F}[U_{\mathrm{obj}}(x_{\mathrm{obj}}, y_{\mathrm{obj}})].
  \end{aligned}
\end{equation}
The optical system can often be abstracted as a finite-space light field modulator, 
which causes the incident light field $U_{\mathrm{in}}$ to undergo amplitude modulation 
$A_{\mathrm{ap}}$
due to finite aperture transmission and phase modulation $\varphi_{\mathrm{ap}}$ caused 
by optical path differences at different spatial positions, 
and spatial mapping effects resulting in an output light field of
\begin{equation}\label{Ap}
  \begin{aligned}
    U_{\mathrm{out}}(x_{\mathrm{opt}}, y_{\mathrm{opt}}) =
    U_{\mathrm{in}}(x_{\mathrm{opt}}, y_{\mathrm{opt}}) \cdot 
    \widetilde{A}_{\mathrm{ap}}(x_{\mathrm{opt}}, y_{\mathrm{opt}}),
  \end{aligned}
\end{equation}
where $\widetilde{A}_{\mathrm{ap}} = {A}_{\mathrm{ap}}\mathrm{exp}[j\varphi_{\mathrm{ap}}]$
is the generalized pupil function.
Finally, the angular spectrum of the output light field $A_{\mathrm{out}}(u, v)$
propagates through a distance of the focal length $f$, forming the 
angular spectrum and light field at the image plane. 
The light intensity detected by the detector is
\begin{equation}\label{FD}
  \begin{aligned}
    I(x, y) = |\mathcal{F}^{-1}[A_{\mathrm{out}}(u, v)
      \cdot \mathrm{exp}(jkf)
      \cdot \mathrm{exp}(-j\pi\lambda f(u^2 + v^2))
    ]|^2.
  \end{aligned}
\end{equation}
Throughout the entire process, the amplitude modulation of the finite aperture 
leads to the constraint of the target information in the frequency domain, 
while the finite area of the detector leads to the constraint of the target 
information in the spatial domain. 
For traditional circular aperture incoherent imaging systems, 
the light field bandwidth product reaching the image plane of the detector is
\begin{equation}\label{SBPc}
  \begin{aligned}
    \mathrm{SBP}_{\mathrm{opt}} = Z\Delta^2\pi / \lambda^2(F/\#)^2.
  \end{aligned}
\end{equation}
Moreover, wave aberration 
causes the complex amplitude of different frequency components in the 
image to be modulated. However, unlike the cut-off of frequency information 
caused by a finite aperture, wave aberration only causes attenuation of information, 
theoretically not leading to information loss.

\subsection{Conversion and Digitization of Optoelectronic Information}\label{2_2}
When the light field reaches the detector, photons will cause the photoelectric 
effect in the photodiode, exciting electrons to form an electrical signal. 
Without considering the Poisson process, the photon number $p$ received 
by the photodetector pixel is determined by its photodetector parameters and 
light information, as
\begin{equation}\label{ps}
  \begin{aligned}
    p = \iint_{\Delta \times \Delta} \frac{\eta(\lambda)\lambda \epsilon_0 tI(x, y)}{2h},
  \end{aligned}
\end{equation}
where $\Delta \times \Delta$ is the pixel area.
The electrical signal induced by the photoelectric effect 
undergoes amplification with gain $K_\mathrm{gain}$ and analog-to-digital 
conversion in the detector circuit, 
forming the discrete digital image as
\begin{equation}\label{DN}
  \begin{aligned}
    \mathrm{DN} = \mathrm{min}\{[K_\mathrm{gain}p], \mathrm{DN}_{\mathrm{sat}}\}
  \end{aligned}
\end{equation}
where $[\cdot]$ is the rounding operation and $\mathrm{DN}_{\mathrm{sat}}$
is the saturation digital number of the optoelectronic detector, 
which typically takes the form of $2^N - 1$. 
In (\ref{ps}), only the intensity of the image field varies with the 
spatial coordinates. Ignoring the case of overexposure, this equation 
can be combined with (\ref{DN}) to express the digital number as
the form of convolution and sampling as
\begin{equation}\label{img}
  \begin{aligned}
    \mathrm{DN}_{\mathrm{img}}(m, n) = 
    k(\lambda) I_{\mathrm{img}}(m\Delta, n\Delta) *
    \mathrm{rect}(m, n) \cdot
    \mathrm{comb}(m, n),
  \end{aligned}
\end{equation}
where $k(\lambda)$ is a spatially independent constant that is related 
only to the detector parameters and the wavelength of light and $x = m\Delta$.
The frequency spectrum of this digital signal is
\begin{equation}\label{imgF}
  \begin{aligned}
    \mathcal{F}[\mathrm{DN}_{\mathrm{img}}(m, n)] = 
    k(\lambda) F_{\mathrm{img}}(u, v) \cdot 
    \mathrm{sinc}(\pi \Delta u, \pi \Delta v) *
    \mathrm{comb}(\Delta u, \Delta v),
  \end{aligned}
\end{equation}
where the term $\mathrm{sinc}(\pi \Delta u, \pi \Delta v)$ represents the decay 
in frequency components caused by window sampling, while the convolution term  
$\mathrm{comb}(\Delta u, \Delta v)$ results in the periodic extension of the 
signal spectrum, causing frequency aliasing when the sampling is insufficient.
From the bandwidth product perspective, when a signal is sampled 
and discretized, its spatial domain and frequency 
domain are coupled. Once the spatial domain characteristics of 
discrete information are fixed, their frequency domain characteristics 
are also fixed.
Moreover, it can be noticed that, 
based on the matching criterion of the optical lens and photodetector 
in remote sensing imaging systems, the optical cut-off frequency 
is located at 1 to 2 times the Nyquist frequency \cite{FNp}, 
which causes the signals at various spatial frequencies 
to be often undersampled, thereby reducing the spatial 
bandwidth product of the discrete information to
\begin{equation}\label{SBPphotod}
  \begin{aligned}
    \mathrm{SBP}_\mathrm{DN} = Z.
  \end{aligned}
\end{equation}
Additionally, due to limitations in detector manufacturing processes, 
the discretization of the light field is also affected by noise. 
Considering Shannon's formula
\begin{equation}\label{ShannonsFormula}
  \begin{aligned}
    C = B\mathrm{log}_2(1 + \mathrm{SNR}),
  \end{aligned}
\end{equation}
where $C$ is the channel capacity, $B$ is the bandwidth, 
and SNR is the signal-to-noise ratio of the channel. 
By applying this formula to spatial imaging scenarios 
and multiplying both sides by the spatial domain width, we obtain
\begin{equation}\label{Information}
  \begin{aligned}
    I_\mathrm{info}(\mathrm{DN}(m, n);I_{\mathrm{obj}}(x, y)) = \mathrm{SBP}\mathrm{log}_2(1 + \mathrm{SNR}),
  \end{aligned}
\end{equation}
where $I$ is the maximal mutual information 
per imaging instance and SBP is the spatial bandwidth 
product of the system.
For example, when the system SNR approaches infinity, 
the signal received by a single detector is an analog 
value that cannot be stored with finite-bit precision. 
Conversely, when quantization noise caused by limited bit 
depth dominates over photoelectric noise, the image information 
content corresponds precisely to the raw file size required 
to store the image.

\subsection{Restoration of Digital Information}\label{2_3}

In information processing systems, methods for image enhancement 
are typically categorized into 3 classes: 1) Physical model-based methods, 
which model system degradation and solve inversely to reconstruct images, 
such as inverse deconvolution reconstruction \cite{NBDB}; 
2) Information enhancement-based methods, which 
heuristically amplify effective components of images, 
such as edge sharpening and filter-based denoising \cite{GIF}; 
3) Learning-based methods, which enhance images by incorporating 
mega data priors, such as generative adversarial networks \cite{DLDB}. 
Compared to the latter two, the restoration approach of the former 
relies on system parameters to maximally preserve real information.
This type of method models the optoelectronic imaging model as
\begin{equation}\label{ImageModel}
  \begin{aligned}
    F_{\mathrm{img}}(u, v) = F_{\mathrm{obj}}(u, v)H(u, v) + N(u, v).
  \end{aligned}
\end{equation}
It integrates the degradation of the system into a degradation kernel 
$H(u, v)$ and integrates photoelectric noise into degradation noise $N(u, v)$, 
meaning that images obtained through photoelectric imaging systems are 
often blurred and noisy. The inverse model of (\ref{ImageModel}) is
\begin{equation}\label{DeconvModel}
  \begin{aligned}
    I_{\mathrm{rec}}(x, y) = \mathcal F^{-1}[
      \frac{F_{\mathrm{img}}(u, v) - N(u, v)}{H(u, v)}
    ],
  \end{aligned}
\end{equation}
where $I_{\mathrm{rec}}$
is the reconstruction information corresponding to objects based on the image.
The spatial domain form $I_{\mathrm{img}}(x, y) - n(x, y)$
of the term $F_{\mathrm{img}}(u, v) - N(u, v)$ in this equation 
is essentially a severely ill-posed denoising problem, 
where shot noise caused by the quantum property 
of photons \cite{Poisson} dominates the photoelectric noise.
Even if photoelectric parameters can be precisely calibrated, 
the noise impact cannot be eliminated.
Under a naive inverse filtering model, both the effective signal and 
noise are amplified proportionally, thus their information content 
remains unchanged.
However, calibration or estimation of the system degradation kernel $H(u, v)$
inevitably contains errors, leading to erroneous modulation in signal 
reconstruction, which in turn generates artifacts degrading image quality, 
such as ringing effects \cite{Ringing}.
On one hand, this causes certain textures to be abnormally amplified, 
particularly in frequency bands with low MTF. 
If these textures are regarded as residual noise after restoration, 
then the information content of the signal is reduced after restoration.
On the other hand, to suppress the abnormal amplification of textures, 
some regularization terms, such as high-frequency suppression \cite{Wiener}
and local smoothing \cite{Laplacian}, are proposed, trading computational 
cost and loss of information content for mitigating the 
signal-to-noise ratio degradation due to abnormal amplification.
\par
In summary, the entire information transmission chain adheres 
to the principle of data processing inequality, meaning that information 
progressively diminishes throughout processing stages as 
\begin{equation}\label{InfoProcess}
  \begin{aligned}
    H_\mathrm{info}[I_{\mathrm{rec}}(x, y)] 
    \le H_\mathrm{info}[I_{\mathrm{img}}(x, y)]
    \le H_\mathrm{info}[I_{\mathrm{obj}}(x, y)].
  \end{aligned}
\end{equation}
Consequently, the information obtainable by photoelectric 
systems constitutes the upper bound of the restored information

The task of the holistic information transmission chain 
can be decoupled as follows: 
1) The photoelectric imaging hardware system aims to 
maximize the preservation of light field information content; 
2) The image restoration algorithm system aims to 
recover target light field information as completely as possible, 
leveraging the photoelectric imaging system parameters and 
acquired information.

\section{Design Methodology based on Holistic Information Analysis}\label{3}

In previous studies, a qualitative principle indicates that for 
frequencies within the optical cut-off, higher MTF values result in 
smaller restoration errors under identical noise levels 
and kernel error spectra \cite{p1}. Therefore, for frequency bands of 
interest, systems should be designed with MTF exceeding a certain 
threshold to counteract restoration errors induced by 
kernel measurement errors, while maximizing SNR to mitigate noise 
amplification during restoration. The conclusion 
that improving MTF and SNR enhances system performance is evident \cite{CNES1, CNES2}.
In this section, we establish a quantitative functional relationship 
demonstrating that the information transmittable by photoelectric 
imaging systems and the restoration accuracy of process systems
increases monotonically with the product of MTF and SNR.

\subsection{MTF-SNR Product Criteria: Imaging Systems}\label{3_1}
To establish MTF and SNR design criteria, begin by analyzing 
information transfer within the hardware system. (\ref{Information}) 
assumes channel corruption by Gaussian white noise, while in 
photoelectric imaging systems with the systemic noise 
such as nonuniform response being well-calibrated, 
the channel is predominantly contaminated by Poisson noise. 
Under sufficient illumination, the instantaneous photon count 
reaching a specific pixel region on the photodetector array approximates \cite{PG}
\begin{equation}\label{PhotonG}
  \begin{aligned}
    p(x, y) + n_p(x, y) \sim \mathcal N(p(x, y), p(x, y)),
  \end{aligned}
\end{equation}
where $n_p(x, y)$ is shot noise caused by light quantum fluctuations
and $\mathcal N$ is the Gaussian distribution.
The statistical characteristics of the shot noise spectrum $N_p(u, v)$ are
\begin{equation}\label{PhotonSC}
  \begin{aligned}
    E[N_p(u, v)] = 0, D[N_p(u, v)] = \sum_{(x, y) \in X \times Y} p(x, y).
  \end{aligned}
\end{equation}
Therefore, the shot noise can be approximated as white noise 
influenced by the entire signal. Furthermore, the spectrum of shot noise, 
formed by the linear sum of Gaussian noises obeying different
means and variances, follows a complex Gaussian distribution. 
Assuming the signal spectrum is also complex Gaussian distributed, 
it satisfies the Gaussian channel information capacity formulation. 
However, the information capacity of individual frequency band 
channels varies with overall signal fluctuations. 
To investigate information transfer in the photoelectric process, 
consider the signal power spectrum as
\begin{equation}\label{SPS}
  \begin{aligned}
    E[|F(u, v)|^2] = |\sum_{(x, y) \in X \times Y} \mathrm{exp}[-j2\pi(ux + vy)]p(x, y)|^2,
  \end{aligned}
\end{equation}
which makes the SNR in each spectrum be
\begin{equation}\label{SNRF}
  \begin{aligned}
    \mathrm{SNR}(u, v) = \sqrt{
      \frac{|\sum_{(x, y) \in X \times Y} \mathrm{exp}[-j2\pi(ux + vy)]p(x, y)|^2}
      {\sum_{(x, y) \in X \times Y} p(x, y)}.
      }
  \end{aligned}
\end{equation}
When $p$ increases by a factor of $m$, the SNR across all 
frequency bands increases by the factor $\sqrt{m}$. 
This equally applies to the digital signal obtained 
after light quanta pass through gain $K$. Furthermore, the SNR in different 
frequency bands depends on the signal itself, and the optical imaging 
channel modulates the complex amplitude of the signal. 
Consequently, at the entire system level, there is
\begin{equation}\label{Information4Image}
  \begin{aligned}
    I_\mathrm{info}(I_{\mathrm{img}}(x, y);I_{\mathrm{obj}}(x, y)) = 
    \sum_{(u, v) \in F_u \times F_v} \mathrm{log}_2(1 + \mathrm{MTF}(u, v)\cdot \mathrm{SNR}(u, v)),
  \end{aligned}
\end{equation}
where $X \times Y$ is the set of spatial position coordinates, 
$F_u \times F_v$ is the set of spatial frequency coordinates, and
\begin{equation}\label{set}
  \begin{aligned}
    |X \times Y| = |F_u \times F_v| = Z.
  \end{aligned}
\end{equation}
It integrates the frequency response capability and 
signal fidelity capability of the imaging system
to characterize the information transfer capacity of the imaging channel. 
In previous research, the MTF-SNR Product (MSP) factor was typically 
used as a heuristic design metric \cite{FNp}, but it is fundamentally 
directly correlated with the information transfer capacity of 
the imaging channel.
\par
In photoelectric imaging hardware systems, consistent MSP can 
achieve equivalent information. This implies that when design objectives 
for a spatial system are fixed particularly in high-resolution spatial 
systems requiring large optical aperture specifications, 
one may enhance the SNR through increased integration stages or 
satellite attitude adjustments for staring imaging \cite{staring}, 
thereby relaxing demands on optical system performance. 
\par
Approximately, for a high-performance baseline system and 
an optimized system, based on (\ref{Information4Image}) and the conclusion 
that SNR is frequency-domain independent, the system should elevate SNR to 
\begin{equation}\label{Ccons}
  \begin{aligned}
    \mathrm{SNR} \approx \mathrm{SNR}_{\mathrm{base}}[\prod_{(u, v) \in F_u \times F_v}
    \frac{
      \mathrm{MTF}_{\mathrm{base}}(u, v)
    }{
      \mathrm{MTF}(u, v)
    }]^{1/Z}
  \end{aligned}
\end{equation}
or employ parameter scanning to compute numerical solutions for implicit equation 
(\ref{Information4Image}) to obtain more precise SNR values.

\subsection{MTF-SNR Product Criteria: Information Restoration}\label{3_2}
At the system level, a consistent MSP level ensures that the upper 
bound of information obtainable by the system remains consistent. 
However, consistent channel capacity does not guarantee consistent information 
acquisition because this also depends on the information input, 
so the fidelity of information transmission is a critical performance 
attribute of the system. From a computational restoration perspective, 
for a specific MSP level and disregarding core errors, inverse 
filtering restoration achieves complete preservation of information 
and reaches the upper bound of the information inequality (\ref{InfoProcess}),
because the SNR and SBP remain unchanged. Set the spectrum of white noise to 
$N_f$ across all frequency bands. When inverse filtering restoration is 
applied, the absolute mean square error (MSE) of the information is given by 
\begin{equation}\label{AMSE}
  \begin{aligned}
    |\varepsilon|^2 = \frac{|N_f|^2}{Z}\sum_{(u, v) \in F_u \times F_v}
    \frac{1}{\mathrm{MTF}^2(u, v)}.
  \end{aligned}
\end{equation}
Clearly, the noise power magnification (NPM) can be defined as
\begin{equation}\label{NPM}
  \begin{aligned}
    \mathrm{NPM} = \frac{1}{Z}\sum_{(u, v) \in F_u \times F_v}
    \frac{1}{\mathrm{MTF}^2(u, v)},
  \end{aligned}
\end{equation}
which characterizes the amplification ratio of noise power during reconstruction. 
For the baseline system and the optimized system, to ensure the consistency of 
information restoration performance, 
\begin{equation}\label{Rcons}
  \begin{aligned}
    \mathrm{SNR} / \sqrt{\mathrm{NPM}} = 
    \mathrm{SNR}_{\mathrm{base}} / \sqrt{\mathrm{NPM}_{\mathrm{base}}}
  \end{aligned}
\end{equation}
should be held. It differs from the form of (\ref{Ccons}) 
because (\ref{Ccons}) characterizes the maximum transmittable information 
of the optical system whereas (\ref{Rcons}) characterizes the actual information 
transmitted by the optical system in an imaging task. These respectively 
represent task-agnostic system channel capability and task-dependent target 
information.The system with high channel capability preserves more 
raw physical features potentially irrelevant to the current task, 
which can provide foundational support for downstream tasks such as 
super-resolution algorithms \cite{SR}.
\par
In image restoration tasks, based on different prior assumptions, 
the restoration process maximizes the corresponding objective function 
at the expense of physical information. For instance in Wiener filtering 
with an additional low-pass filter, the absolute MSE of the information is 
given by 
\begin{equation}\label{Wiener}
  \begin{aligned}
    |\varepsilon|^2 \approx \frac{1}{Z}(\sum_{(u, v) \in F_u \times F_v} \frac{|N_f|^2Lp^2(u, v)} 
    {\mathrm{MTF}^2(u, v)} +
    2\sum_{(u, v) \in F_u \times F_v} F_{\mathrm{obj}}(u, v)\overline{Lp}(u, v) \times
    \sum_{(u, v) \in F_u \times F_v} \frac{|N_f|Lp(u, v)}{\mathrm{MTF}(u, v)}),
  \end{aligned}
\end{equation}
where
\begin{equation}\label{Lp}
  \begin{aligned}
    Lp(u, v) = \frac{1}{1 + 1 / [\mathrm{MTF}(u, v) \cdot \mathrm{SNR}(u, v)]^2}.
  \end{aligned}
\end{equation}
As the frequency band increases, the reduction in the MTF value causes 
the enhancement of low-pass filtering. Compared to (\ref{AMSE}), 
the first term of (\ref{Wiener}) adds a noise suppression term from low-pass 
filtering but relatively the latter term causes damage to 
the information. In fact, all restoration formulas possessing 
the form of 
\begin{equation}\label{Restoration}
  \begin{aligned}
    F_\mathrm{rec} = \mathrm{argmin}_{F_\mathrm{obj}}||HF_\mathrm{obj} - F_\mathrm{img}||_2^2
    + \lambda P(F_\mathrm{obj})
  \end{aligned}
\end{equation}
have consistent restored MSE form expressed in the form of 
\begin{equation}\label{RestorationMSE}
  \begin{aligned}
    |\varepsilon|^2 \approx \frac{1}{Z}(\sum_{(u, v) \in F_u \times F_v} \frac{|N_f|^2} 
    {\mathrm{MTF}^2(u, v)} + 
    2\frac{\lambda}{\mathrm{MTF}^2(u, v)}
    \frac{\partial P(F_\mathrm{obj})}{\partial F_\mathrm{obj}}|_{F_\mathrm{obj} = \hat F_\mathrm{obj}}
    \sum_{(u, v) \in F_u \times F_v} \frac{|N_f|}{\mathrm{MTF}(u, v)}),
  \end{aligned}
\end{equation}
where $H$ is the optical transfer function (OTF) and $P$ is
the prior function of the image.
So based on the specific task requiring information, 
a trade-off should be made between the maximum obtainable information 
and the fidelity of the information.

\subsection{System Design Based on MTF-SNR Product Criteria}\label{3_3}

Based on the analysis of holistic information in previous sections, 
we can abstract a system design to 
\begin{equation}\label{SystemDesign}
  \begin{aligned}
    \mathrm{max}\ &\lambda_I I + \lambda_E E \\
    s.t.\ &I = \sum_{(u, v) \in F_u \times F_v} \mathrm{log}_2(1 + \mathrm{MTF}(u, v)\cdot \mathrm{SNR}(u, v)), \\
    & E = -\sum_{(u, v) \in F_u \times F_v} |\varepsilon(u, v)|^2
  \end{aligned}
\end{equation}
where $\lambda_I$ and $\lambda_E$ should be determined by the specific 
task to favor signals containing more true information or signals more 
similar to the original signal. Considering engineering requirements,
when requirements are specified, there should be
\begin{equation}\label{ImageSystemDesign}
  \begin{aligned}
    \mathrm{min}\ &\lambda_\mathrm{MTF} \mathrm{MTF} +  \lambda_\mathrm{SNR} \mathrm{SNR}\\
    s.t.\ &E \ge E_\mathrm{base},
  \end{aligned}
\end{equation}
where $\mathrm{MTF}$ and $\mathrm{SNR}$ are determined by the system while 
$\lambda_\mathrm{MTF}$ and $\lambda_\mathrm{SNR}$ are determined by the cost 
of improving the MTF and SNR respectively. It is often not formalized in design 
processes but it embodies the core design concept. When designing a space 
system with the smallest possible aperture, there should be $\lambda_\mathrm{SNR} \ll
\lambda_\mathrm{MTF}$. In other words,
under the imaging and restoration requirements, the reduction in MTF can be 
compensated by improving SNR to complete the system design.

\begin{table}[htbp]
  \caption{Main space system parameters}\label{table1}
  \centering
  \begin{tabular}{ccc}
  \toprule
  Parameter & Symbol & Value \\ 
  \midrule
   Orbit altitude & $H_\mathrm{orbit}$ & 500 km \\
    Orbit inclination & $i_\mathrm{orbit}$ & 98$^\circ$ sun synchronous \\
   \midrule
    Atmospheric model & - & Mid-latitude summer \\ 
   \midrule
    Pixel size & $\Delta$ & 10 $\mu$m \\
    Pixel number & $Z$ & 4096 $\times$ 4096 \\
    TDI stage & $M$ & 8$\sim$128 \\
    Full well  & $p_\mathrm{FW}$ & 250 ke- \\
    Quantum efficiency & $\eta$ & about 50\% \\
    Line frequency & $f_L$ & 30 kHz \\
    Quantization bits & - & 12 bit \\
    Spectral band & $\lambda$ & 400 nm $\sim$ 900 nm \\
  \bottomrule
  \end{tabular}
  \end{table}

\section{Design Case based on Holistic Information Theory}\label{4}

\subsection{Benchmark System}\label{4_1}
The primary technology for high-resolution optical remote sensing imaging remains 
large-aperture monolithic optical remote sensing imaging systems. 
Considering aperture design for a space optical system on a small satellite, 
the satellite platform, atmospheric environment, and photodetector parameters 
adopt reasonable values as shown in Table \ref{table1} \cite{SatelliteParameters}. 
Under these parameters, 
the design target achieves a ground sampling distance (GSD) of 0.1 m in the visible band. 
This determines the optical focal length $F$ as 50 m and the 
Nyquist frequency $f_\mathrm{nyq}$ as 50 lp/mm. 
If the catadioptric configuration is selected for the space optical system and 
diffraction-limited performance is achievable, this indicates the 
pupil function is a circular amplitude modulation function. 
Based on the central wavelength and 
conventional design specifications of $\mathrm{MTF@Nyq} \ge 0.3$, the optical aperture 
is determined to be approximately 2800 mm.
When the system receives sufficient uniform radiation 
at 0° solar zenith angle and 0° satellite zenith angle, 
the SNR reaches about 100 at $\rho = 0.2$ using 16-stage integration \cite{SNRParameters},
where $\rho$ denotes ground reflectance, assigned a common value of 0.2 \cite{RParameters}. 
Without loss of generality, inverse filtering restoration is applied for 
fundamental system analysis. 
Based on (\ref{NPM}), the NPM of this system can be calculated as 8.56, 
meaning that after image reconstruction, the noise level will be amplified 
to approximately 3 times.

\begin{figure}[t]\rmfamily
\centering
\includegraphics[width=0.45\textwidth]{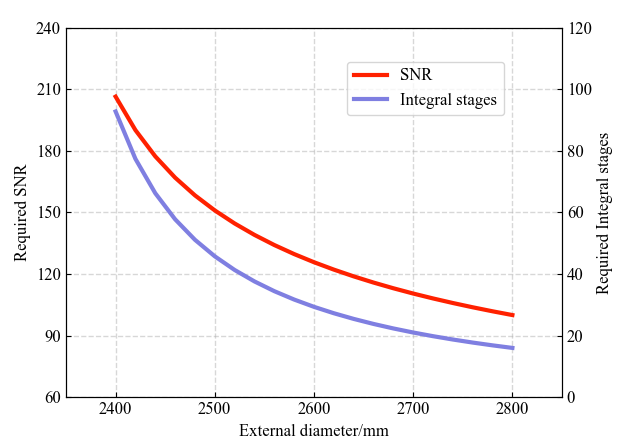}
\caption{
  The required SNR and integration stages vary with optical aperture to 
  maintain an equal level of MSE in the reconstructed image.
} 
\label{Fig1} 
\end{figure}

\begin{figure*}[t]\rmfamily
\centering
\includegraphics[width=0.925\textwidth]{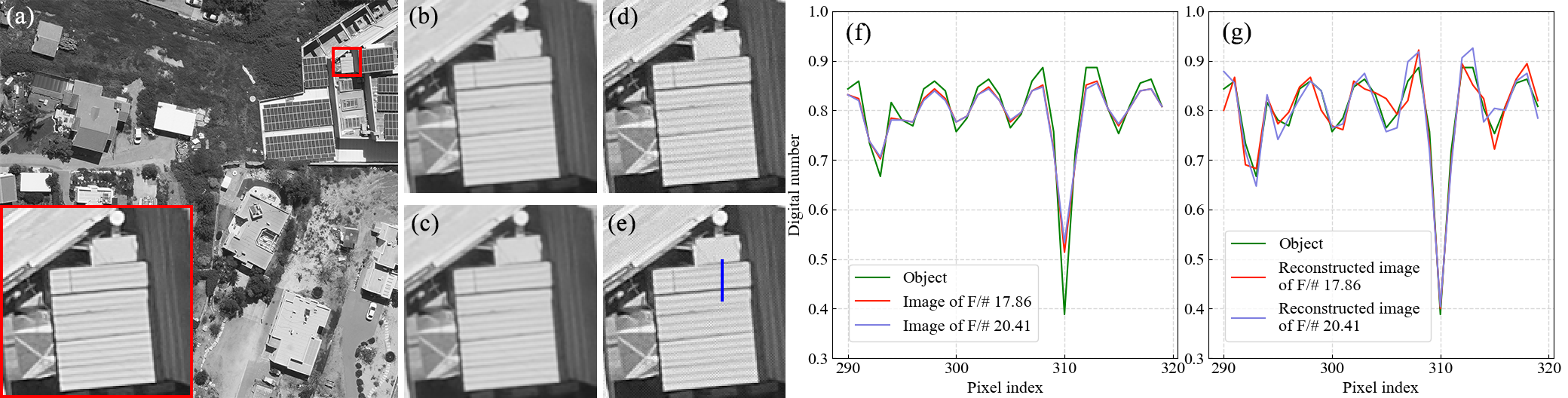}
\caption{
  Simulated imaging and computational reconstruction results of the same 
  target object under a high-MTF, low-SNR system and a low-MTF, high-SNR 
  system. The target object used in the simulation experiment 
  and the region of interest labeled by red box is shown in (a). 
  The images of the target region with the two systems
  is shown in (b) and (c). 
  The reconstructed images based on system parameters and the images
  is presented in (d) and (e).  The grayscale values of periodic 
  textures labeled by the blue line 
  within the target regions is illustrated in (f) and (g).
  The system with a higher F-number produces images with lower contrast 
  of the target texture, but after reconstruction, both images 
  maintain consistent contrast and fidelity.
} 
\label{Fig2} 
\end{figure*}

\begin{figure*}[t]\rmfamily
\centering
\includegraphics[width=0.925\textwidth]{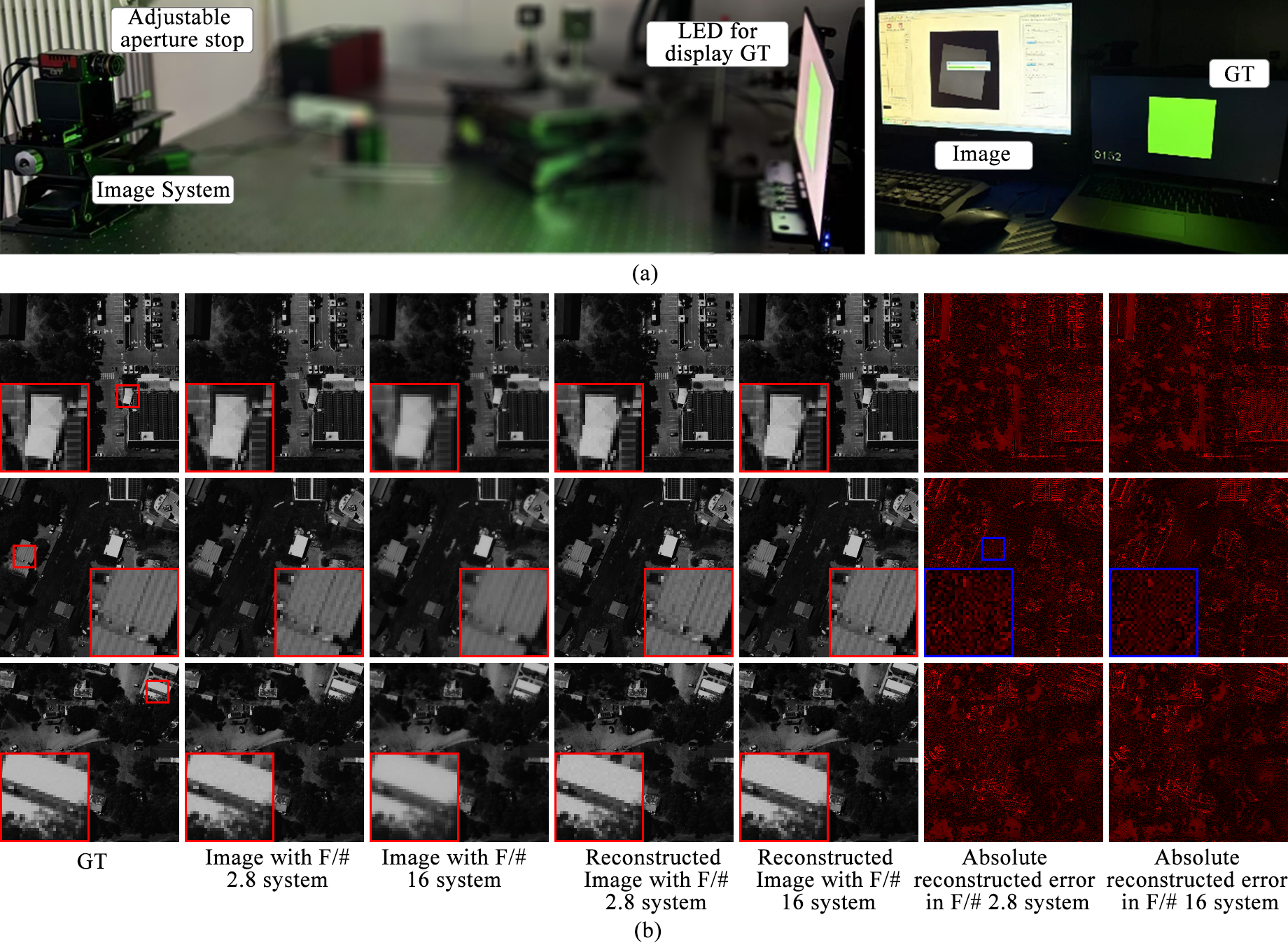}
\caption{
Experimental setup and results. Figure (a) shows the experimental setup, where 
the grayscale-to-radiance mapping relationship for the LED and the non-uniform 
response of the camera are calibrated as much as possible. 
Figure (b) shows the experimental results, examining the PSNR of reconstructed 
images for three targets, The PSNR values are 49.91, 50.50, and 50.37
in the F/\# 2.8 system, respectively, while there are the values of 50.42, 50.67, and 50.92
in the F/\# 16 system. The reconstruction quality of both systems was basically maintained at an 
equivalent level. Please zoom in to observe pixel level details.
} 
\label{Fig3} 
\end{figure*}

\subsection{Design of Small Aperture Imaging System}\label{4_2}
In Section 4.1, to achieve high-resolution remote sensing, 
its optical aperture needs to reach 2800 mm, leading to a large mechanical envelope size. 
The mass of the optical system often increases with the cube of its outer diameter \cite{Principles}, 
greatly increasing the deployment cost of the satellite due to its size and mass,
which poses a challenge for the construction of large-scale constellations. 
Reducing the primary mirror size is the most direct means to reduce engineering costs, 
therefore we analyze based on the mathematical model proposed in Section 3.3 for the requirements 
in Section 4.1. \par

In Section 3, we demonstrated that the MSP is the decisive factor affecting information acquisition 
and restoration in imaging systems. This suggests that we can enhance tolerance to image quality 
degradation by improving the SNR. Under sufficient radiation, the factors affecting 
SNR in imaging systems are the area of the optical aperture and the integration time, 
respectively by increasing the area receiving radiation and the time to increase the photons
received by the detector as
\begin{equation}\label{PhotonNum}
  \begin{aligned}
    \mathrm{SNR} = \frac{\Delta\sqrt{
      \pi\int_{\lambda_\mathrm{min}}^{\lambda_\mathrm{max}}\eta(\lambda)(\lambda/hc)(M/f_L) L(\lambda)
      \tau(\lambda)d\lambda
    }}{2F/\#},
  \end{aligned}
\end{equation}
where $L(\lambda)$ and $\tau(\lambda)$ 
denote the irradiance in the entrance
aperture and the transmittance of the optical system at a specific wavelength respectively. 
By adjusting the integration stages of the detector, the degradation of MTF and SNR due to 
aperture reduction can be compensated. Therefore, we gradually reduced the aperture of the optical 
system and observed the changes in the required integration stages to maintain the same level 
of reconstruction error, as shown in Fig. 1. When electronic pressure is low, integration time 
can greatly alleviate the demand for the optical aperture. For example, using 64 integration stages 
allows reducing the optical aperture to 2450 mm, with the envelope size and mass of the 
primary mirror decreasing by approximately 20\% and 30\% respectively based on \cite{Principles}.

\subsection{Simulation and Imaging Experiments}\label{4_3}
To verify the theoretical correctness of the design in Section 4.2, 
we conducted a simulation experiment. We selected a high-definition 
image as the target object, shown in Fig. 2(a). 
First, using the parameters from Section 4.2, we applied optical 
degradation under paraxial approximation with F/\# values of 17.86 and 20.41, 
and added Poisson noise at SNR levels of 100 and 170 under saturated grayscale
to give images of the target object. 
We focused on a texture-rich region 
and the images are shown in Fig. 2(b) and 2(c). Next, using the correct blur kernels, 
we performed inverse filtering reconstruction to get the reconstructed 
images for regions 2(d) and 2(f). The PSNR representing the MSE of 
reconstruction for these two reconstructed images is 
33.79 and 33.68,
showing consistent restoration quality. 
\par
Additionally, we conducted a physical experiment. 
We modified the MTF and SNR by changing the aperture of the optical lens 
and the exposure time of the detector. Here, adjusting the aperture results in 
F-numbers of 2.8 and 16 for the optical lens. The pixel size is 
6.45 $\mu$m, and the center wavelength of the narrowband LED light source is 532 nm. 
We adjusted the exposure times for the two experiments to make the imaging 
grayscale reach half-saturation. Based on (\ref{NPM}), the NPM values for 
the F/\# 2.8 and F/\# 16 systems are 1.28 and 23.64, respectively. 
Then, according to (\ref{Rcons}), the SNR of the F/\# 16 system should 
be about 4.3 times that of the F/\# 2.8 system. 
Therefore, the F/\# 16 shooting is extended to 18 times. 
The entire imaging system was well-calibrated. The experiment and results are 
shown in Fig. 3. 
Each column represents one target. The blur levels of the images are 
very different between the two systems. However, after reconstruction, 
the similarity to GT was basically consistent. In the reconstructed 
image of the F/\# 2.8 system, noise was amplified more uniformly. 
In contrast, in the F/\# 16 system, high-frequency noise 
was amplified more severely, resulting in it shows a grid-like noise texture
as shown in the absolute reconstructed error in Fig. 3(b). 
Additionally, due to residual systematic errors in calibration, there are edge 
textures and grayscale bias in the reconstruction error, 
but this is consistent for both systems and does not affect the conclusion.
\par
In summary, when the imaging and computational restoration system is 
considered as a whole, the MTF and SNR of the imaging system, as limitations 
on information transmission, can compensate for each other to achieve consistent 
computational imaging results. This means that for high-resolution remote sensing 
imaging requirements, the requirements for large-aperture and high-precision lens 
fabrication can be relaxed. Instead, compensation can be achieved through the fine 
design of photoelectric detectors and the staring imaging method to meet 
the set imaging performance targets.

\section{Conclusion and Future Work}\label{5}
This paper proposes the concept of holistic information imaging, 
which jointly considers the optical subsystem, electronic subsystem, 
and computational subsystem of an imaging system. This establishes 
quantitative criteria for collaboration between these subsystems. 
Specifically, for standard monolithic remote sensing imaging systems, 
we demonstrate through theoretical analysis and experimentation that system 
design can relax MTF requirements by improving SNR. This avoids the requirements for 
large-aperture lenses and high-precision surface fabrication. 
This theory has been applied to the development of the Jilin-1 
constellation satellites, enabling ultra-high-resolution Earth 
remote sensing at acceptable manufacturing costs.
\par
However, this work still faces challenges. First, while we analyze the 
information transmission chain in holistic information imaging, 
the advantages of increased information capacity need validation across 
diverse application scenarios. Thus, investigating broader interactions between 
computational algorithms and imaging systems remains a key future task. 
Additionally, the uncertainty in degradation calibration during imaging is 
an important problem affecting engineering applications.
Finally, wider imaging paradigms like hybrid refractive-diffractive optical 
systems and metalens systems can aslo be treated as information transmission 
systems. Integrating all imaging systems into a unified framework 
constitutes another critical future direction.

\bibliographystyle{unsrtnat}
\bibliography{cas-refs}

\end{document}